\documentclass{sf2a-conf2017}
\usepackage{graphicx}
\usepackage{hyperref}
\usepackage[]{natbib}  
\usepackage{epstopdf}
\usepackage{mathrsfs}
\usepackage{amsbsy}
\usepackage{color}

\def\BibTeX{{\rm B\kern-.05em{\sc i\kern-.025em b}\kern-.08em
    T\kern-.1667em\lower.7ex\hbox{E}\kern-.125emX}}
\bibpunct{(}{)}{;}{a}{}{,}  %%%%%%%%%%%%%  A&A bibliography style
\newcommand{\smc}[1]{#1}
\begin{document}

\TitreGlobal{SF2A 2017}

%%-----------------------------------------------------------------
%%      the top matter
%%

\title{Repercussions of thermal atmospheric tides on the rotation of terrestrial planets in the habitable zone}

\runningtitle{Repercussions of thermal atmospheric tides on the rotation of terrestrial planets}
% Impact of atmospheric stratification on the rotation of terrestrial planets driven by thermal tides

\author{P. Auclair-Desrotour}\address{LAB, Universit\'e de Bordeaux, CNRS UMR 5804, Universit\'e de Bordeaux - Bât. B18N, All\'ee Geoffroy Saint-Hilaire CS50023, 33615 Pessac Cedex, France}

\author{S. Mathis$^{2,}$}
\address{Laboratoire AIM Paris-Saclay, CEA/DRF - CNRS - Universit\'e Paris Diderot, IRFU/DAp Centre de Saclay, F-91191 Gif-sur-Yvette Cedex, France}
\address{LESIA, Observatoire de Paris, PSL Research Universit\'e, CNRS, Sorbonne Universit\'e, UPMC Univ. Paris 6, Univ. Paris Diderot, Sorbonne Paris Cit\'e, 5 place Jules Janssen, F-92195 Meudon, France}

\author{J. Laskar}\address{IMCCE, Observatoire de Paris, CNRS UMR 8028, PSL Research University, 77 Avenue Denfert-Rochereau, 75014 Paris, France}

%% IF Author3 has the same affiliation than Author1:
%\author{C.\,E. Author3$^1$}

%% IF Author3 has its own affiliation:
%\author{C.\,E. Author3}\address{Dept. of Chess, University of Games, 35101 Las Vegas, Monaco} 

%% IF Author3 has two affiliations, the one of Author1 and a second one:
%\author{C.\,E. Author3$^{1,}$}\address{Dept. of Chess, University of Games, 35101 Las Vegas, Monaco} 

%% Keep this line, even if the page will be settled afterwards.
\setcounter{page}{237}

%%-----------------------------------------------------------------

\maketitle

%%-----------------------------------------------------------------
%%        The abstract
%% 
%%  Warning!  within the abstract:
%%  - do not use macros. 
%%  - do not use commands like: \cite, \citet, \citep ... etc.

\begin{abstract}
Semidiurnal atmospheric thermal tides are important for terrestrial exoplanets in the habitable zone of their host stars. With solid tides, they torque these planets, thus contributing to determine their rotation states as well as their climate. Given the complex dynamics of thermal tides, analytical models are essential to understand its dependence on the structure and rotation of planetary atmospheres and the tidal frequency. In this context, the state of the art model proposed in the 60’s by Lindzen and Chapman explains well \smc{the properties of} thermal tides in the asymptotic regime of Earth-like rapid rotators but predicts a non-physical diverging tidal torque in the vicinity of \smc{the} spin-orbit synchronization. In this work, we present a new \smc{model} that addresses this issue by taking into account dissipative processes through a Newtonian cooling. First, we recover the tidal torque recently obtained with numerical simulations using General Circulation Models (GCM). Second, we show that the tidal response is very sensitive to the atmospheric structure, particularly to the stability with respect to convection. A strong stable stratification is able to annihilate the atmospheric tidal torque, leading to synchronization, while a convective atmosphere will be submitted to a strong torque, leading to a non-synchronized \smc{rotation} state.
\end{abstract}

%% Insert the keywords (to appear in the ADS indexing)
%% Keywords must be separated by a comma

%%-----------------------------------------------------------------

\section{Introduction}
%%---------------------

The dynamics of planetary systems over long times scales is tightly related to mutual interactions between their bodies, and particularly tides \smc{\citep[e.g.][]{MR2013}}. Tides \smc{affect} the rotation of planets, their distance to the star, and can generate a strong tidal heating, which is able to modify significantly the surface equilibrium of planets as observed for instance on the Jupiter's satellite Io \citep[][]{Lainey2009}. Therefore, with the continuously growing number of discovered terrestrial planets orbiting in the habitable zone of their host stars, it is today crucial to understand physically the role played by tides \smc{for} planetary dynamics. In this work, we focus on thermal atmospheric tides, namely tides generated by the stellar heating of a planetary atmosphere. These tides are of great importance for terrestrial planets close to spin-orbit synchronization because they are able to torque them \textit{away} from this equilibrium state, in opposition to gravitationally excited tides. Typically, in the Solar system, they contribute to lock Venus at a non-synchronized rotation rate where the solid and atmospheric tidal torques exactly balance each other \citep[][]{GS1969,DI1980,CL2001}. In the absence of atmospheric tides, Venus would be today synchronized owing to the relatively small evolution timescales associated with gravitational tides. Hence, considering both the potential number of Venus-like terrestrial planets among \smc{the exoplanets} discovered during the last decade and the impact that the \smc{rotation} of a planet has on \smc{its} atmospheric general circulation and climate, it is necessary to \smc{understand} the dependence of the atmospheric tidal response on the physical properties of \smc{planetary} atmosphere and the tidal frequency. 

This question was addressed recently in a numerical way by \cite{Leconte2015}, who used a General Circulation Model (GCM) to compute the \smc{atmospheric} tidal torque. However, this approach is limited by the high computational cost of numerical simulations \smc{that do not yet allow to explore a broad domain of the parameters space}. The classical theory of atmospheric tides \citep[see e.g.][]{CL70} offers an elegant analytic approach where the tidal response is treated as a linear perturbation and written as an explicit function of the atmospheric parameters. This approach, which ignores dissipative effects, explains well the Earth's atmospheric tides and more generically the case of rapid rotators. However, it has to be adapted to the case of slowly rotating planets where dissipative processes cannot be neglected any more. Thus, we adopt \smc{in this study} the linear approach and generalize it to slowly rotating planets by introducing a Newtonian cooling. We compute the tidal response and the corresponding semidiurnal tidal torque first using a global modeling, second with a local Cartesian one. We \smc{show} how the atmospheric structure affects the tidal torque through the \smc{stability of} stratification of the fluid with respect to convection. \smc{We provide} here the main lines of results, which are detailed in \cite{ADLM2017a}, \smc{\cite{ADLM2017b}} and \smc{\cite{ADML2017c}}.

\section{Physical setup}

\begin{figure}[ht!]
 \centering
 \includegraphics[width=0.54\textwidth,clip]{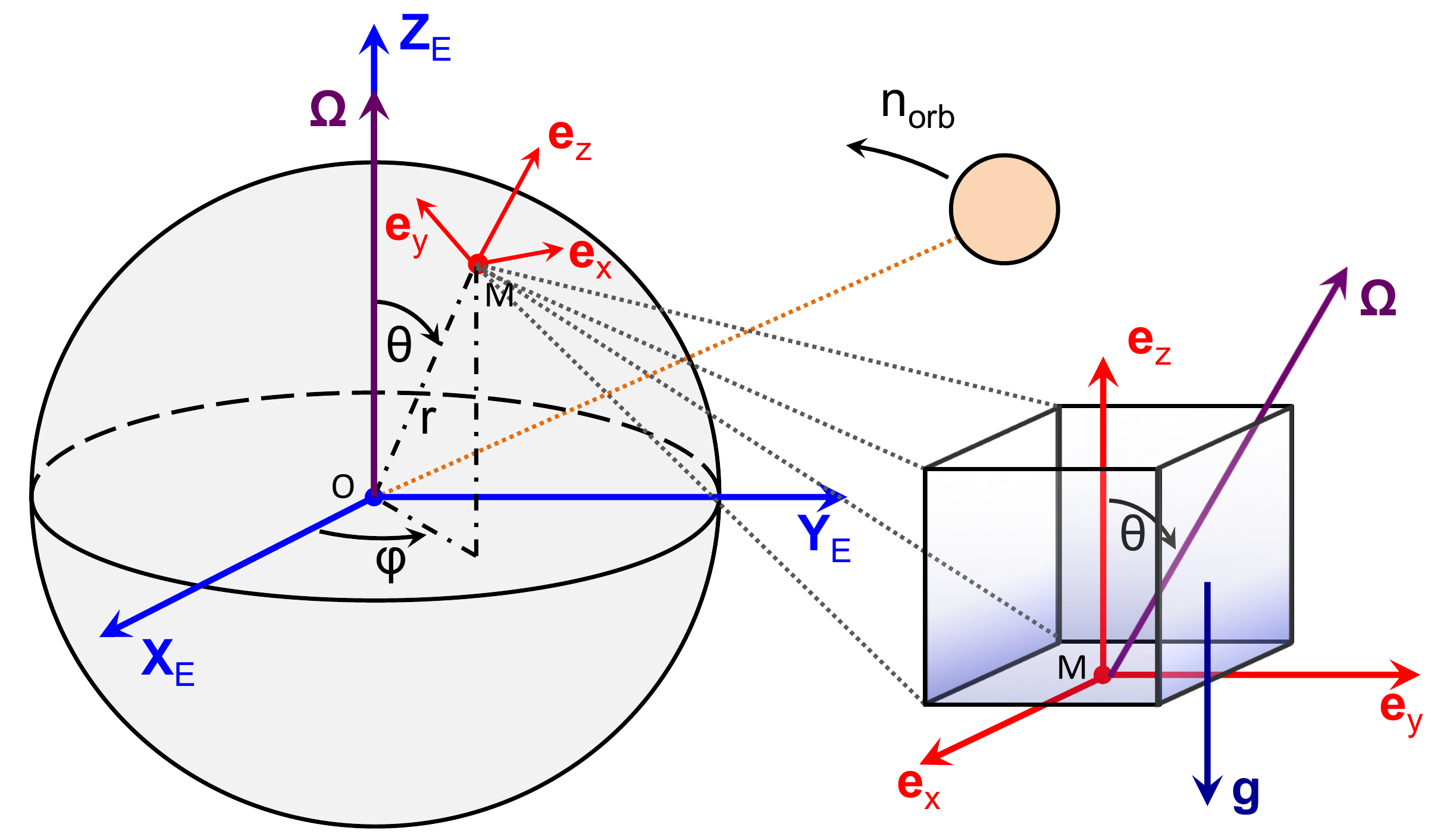} \hspace{0.3cm} 
 \includegraphics[width=0.42\textwidth,clip]{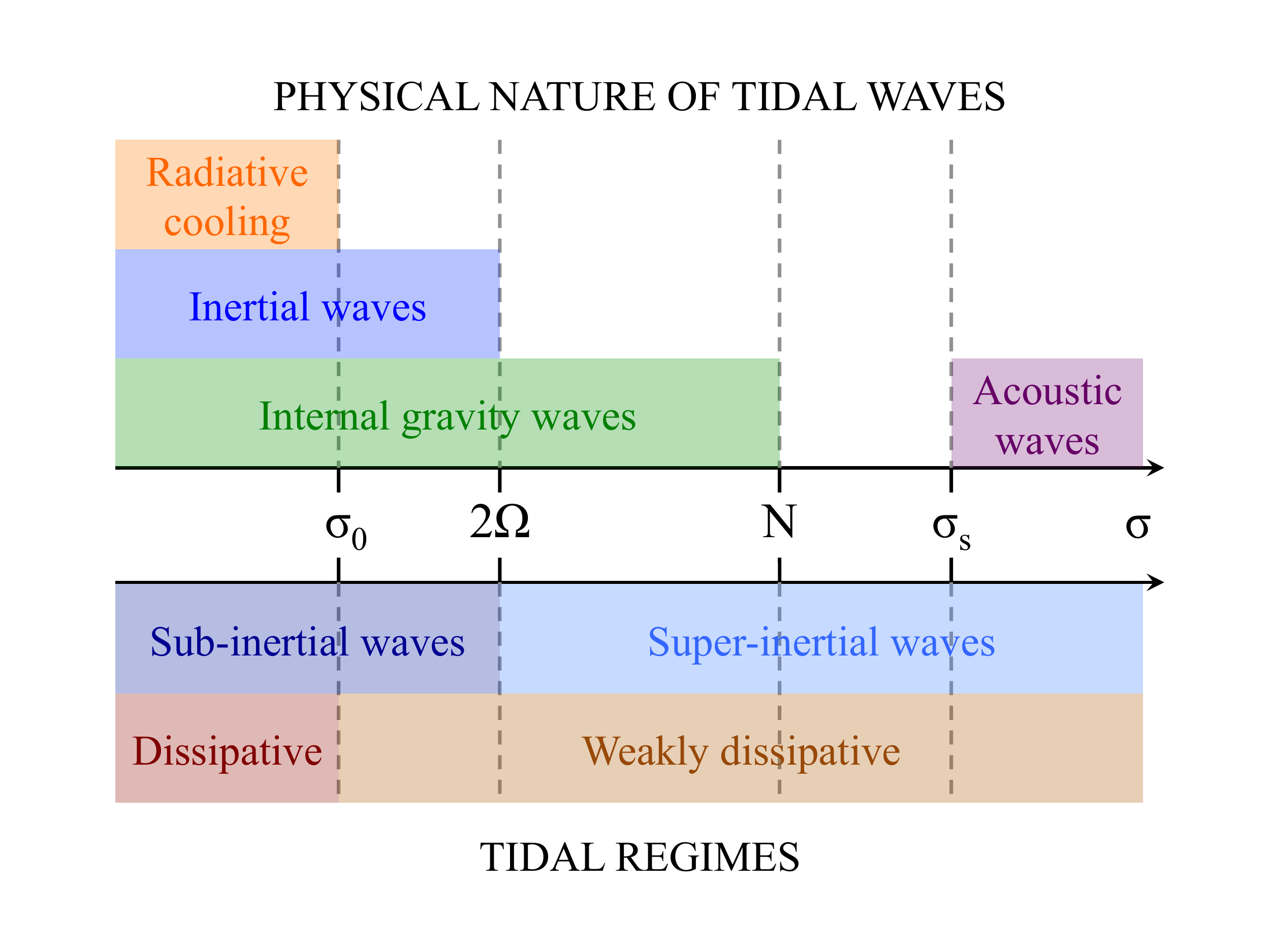}      
%% Note the ABSENCE of the extension .pdf  !
  \caption{{\bf Left:} Spherical and Cartesian reference frames and systems of coordinates. The vectors $\boldsymbol{\Omega}$ and $ \textbf{g} $ designate the rotation and the gravity respectively. The notation $ n_{\rm orb} $ refers to the orbital frequency of the star in the reference frame of the planet. {\bf Right:} Frequency spectrum of tidal regimes and waves characterizing the \smc{atmospheric} tidal response. The parameter $ \sigma $ designates the forcing frequency, $ 2 \Omega $ the inertia frequency, $N$ the Brunt-V\"ais\"al\"a frequency, and $ \sigma_{\rm s}$ the characteristic acoustic cutoff frequency.}
  \label{fig:schema_box}
\end{figure}

We consider a spherical terrestrial planet rotating at the spin angular velocity $ \Omega $ and orbiting its host star at the orbital frequency $ n_{\rm orb} $ (see Fig.~\ref{fig:schema_box}, left panel). In the absence of perturbation, the fluid is supposed to be at the hydrostatic equilibrium and radially stratified (background distributions only depend on the vertical coordinate $ r$). The stratification of the atmosphere with respect to convection is characterized by the Brunt-V\"ais\"al\"a frequency given by

\begin{equation}
N^2 = - g \left[ \frac{d \ln \rho_0}{dr} - \frac{1}{\Gamma_1} \frac{d \ln p_0}{dr} \right],
\end{equation}

\noindent where $ g $ designates the gravity, $ \Gamma_1 $ the adiabatic exponent, and $ p_0 $ and $ \rho_0 $ pressure and density background distributions respectively. Taking into account radiative losses through a Newtonian cooling introduces an additional frequency $ \sigma_0 $, which is the inverse of the thermal timescale. Finally, owing to the compressibility of the fluid, the gravest horizontally propagating Lamb modes can be excited if the forcing frequency $ \sigma $ is greater than the characteristic acoustic cutoff frequency $ \sigma_{\rm s} $. Hence, the hierarchy of the characteristic frequencies $ \sigma_0 $, $ 2 \Omega $, $ N $, $ \sigma_{\rm s} $ and $ \sigma $ fully determines the nature of the tidal response and which kind of waves it is composed of (see Fig.~\ref{fig:schema_box}, right panel): \smc{i.e.} inertial, gravity and acoustic waves damped by radiative cooling. 

\newpage

\smc{Two types of models are used:}
\begin{itemize}
 \item[$\bullet$] Global model \smc{in} spherical geometry. \smc{The background atmospheric structure only depends on the radial coordinate}. \smc{We assume} the \emph{traditional approximation}, \smc{which filters out the components of the Coriolis acceleration related to the latitudinal projection of the rotation vector} \citep[\smc{we refer the reader to}][\smc{for a complete discussion}]{ADLM2017a}. 
 \item[$\bullet$] Local model \smc{constituted by a} Cartesian local fluid section of the atmosphere (Fig.~\ref{fig:schema_box}, left panel) \smc{\citep[][]{ADML2017c}}. \smc{we assume the} \emph{f-plane} and \emph{anelastic approximations} \smc{(where acoustic waves are filtered out). In this model, the complete Coriolis acceleration is taken into account.} 
\end{itemize}

%%
%% Example of single figure
%%

\section{Thermally generated atmospheric tidal torque}

\subsection{Global model}

\begin{figure}[ht!]
 \centering
 \includegraphics[width=0.6\textwidth,clip]{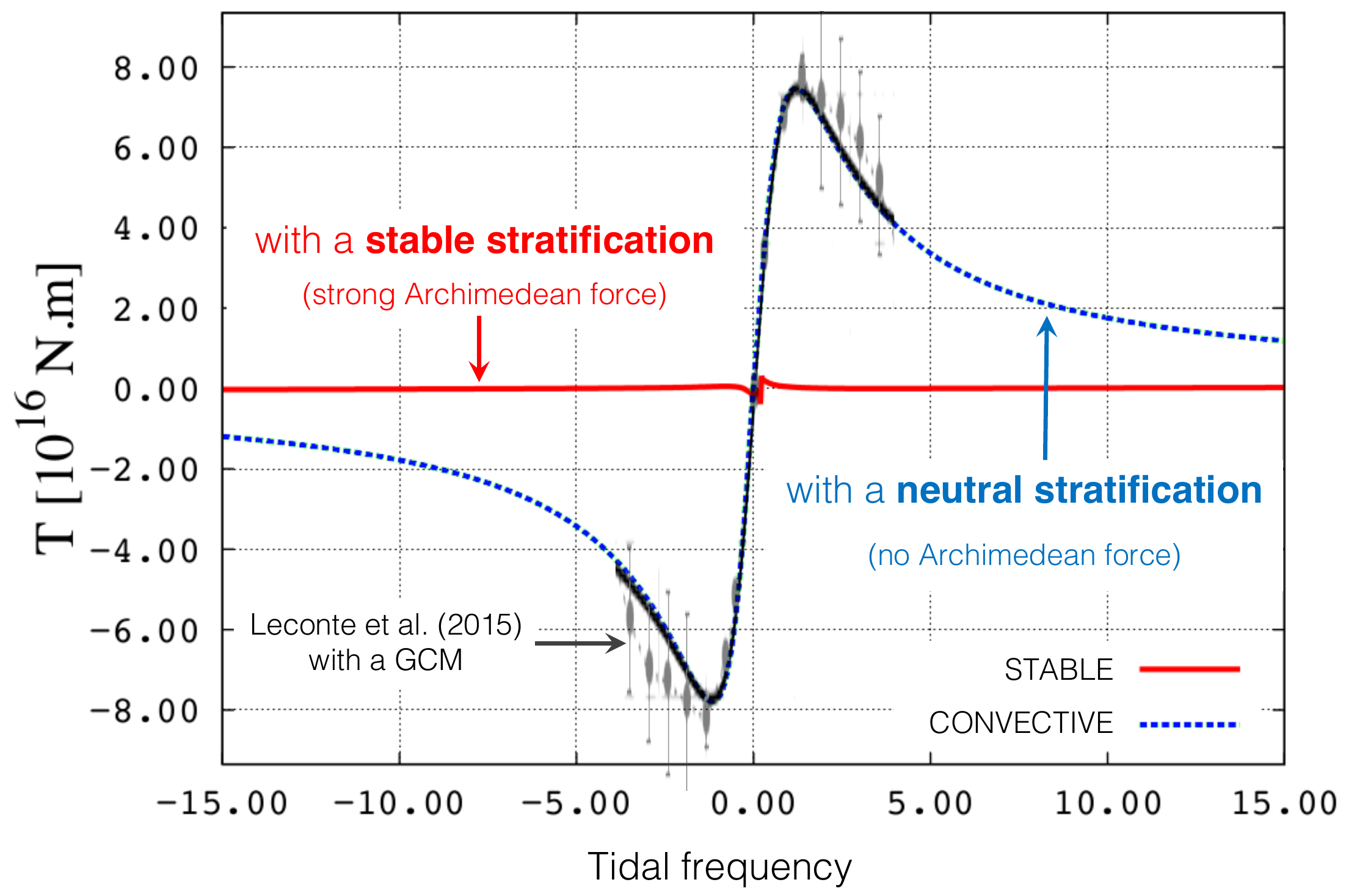}      
%% Note the ABSENCE of the extension .pdf  !
  \caption{Tidal torque given by the global model as a function of the tidal frequency $ \omega = \left( \Omega - n_{\rm orb} \right)  $ for two different atmospheric structure: stably-stratified (red solid line) and neutrally-stratified (blue solid line) with respect to convection. For comparison, the shape of the tidal torque obtained by \cite{Leconte2015} using a GCM is plotted in addition to the two previous cases. Grey dots \smc{correspond} to \smc{numerical} simulations, the black solid line to a fit of these results with a Maxwell model.}
  \label{fig:torque}
\end{figure}

In \cite{ADLM2017a}, we compute for two different atmospheric structures the tidal torque exerted on the atmosphere caused by the quadrupolar component of the thermally excited semidiurnal tide. The first structure corresponds to an isothermal atmosphere, where $ N \gg \sigma $. It is strongly stably stratified with respect to convection. The second structure is an \smc{isentropic} profile of temperature, which is such that $ N = 0 $ and thus enforces a neutral stratification. The tidal torques that we obtain in these two cases are plotted on Fig.~\ref{fig:torque} as functions of the normalized tidal frequency $ \omega = \left( \Omega - n_{\rm orb} \right) / n_{\rm orb} $ using for parameters a set of values derived from the Venus case. As can be seen, we recover \smc{in the neutral stratification case} the tidal torque obtained by \cite{Leconte2015} with GCM simulations \smc{for Venus}. The observed behaviour corresponds to that described by the Maxwell model: a global tidal bulge follows the perturber with a delay depending of the thermal inertia of the atmosphere. In the other case, the strong stable stratification of the atmosphere prevents a net tidal bulge to form, a local density decrease being immediately compensated by the motion of denser fluid particles driven upward by the Archimedean force. As a consequence, the tidal torque is very weak and cannot balance the solid torque as in the \smc{neutral} case. 

\subsection{Local model}

\begin{figure}[ht!]
 \centering
 \includegraphics[height=0.36\textwidth,trim = 1.5cm 1.6cm 6cm 1.5cm,clip]{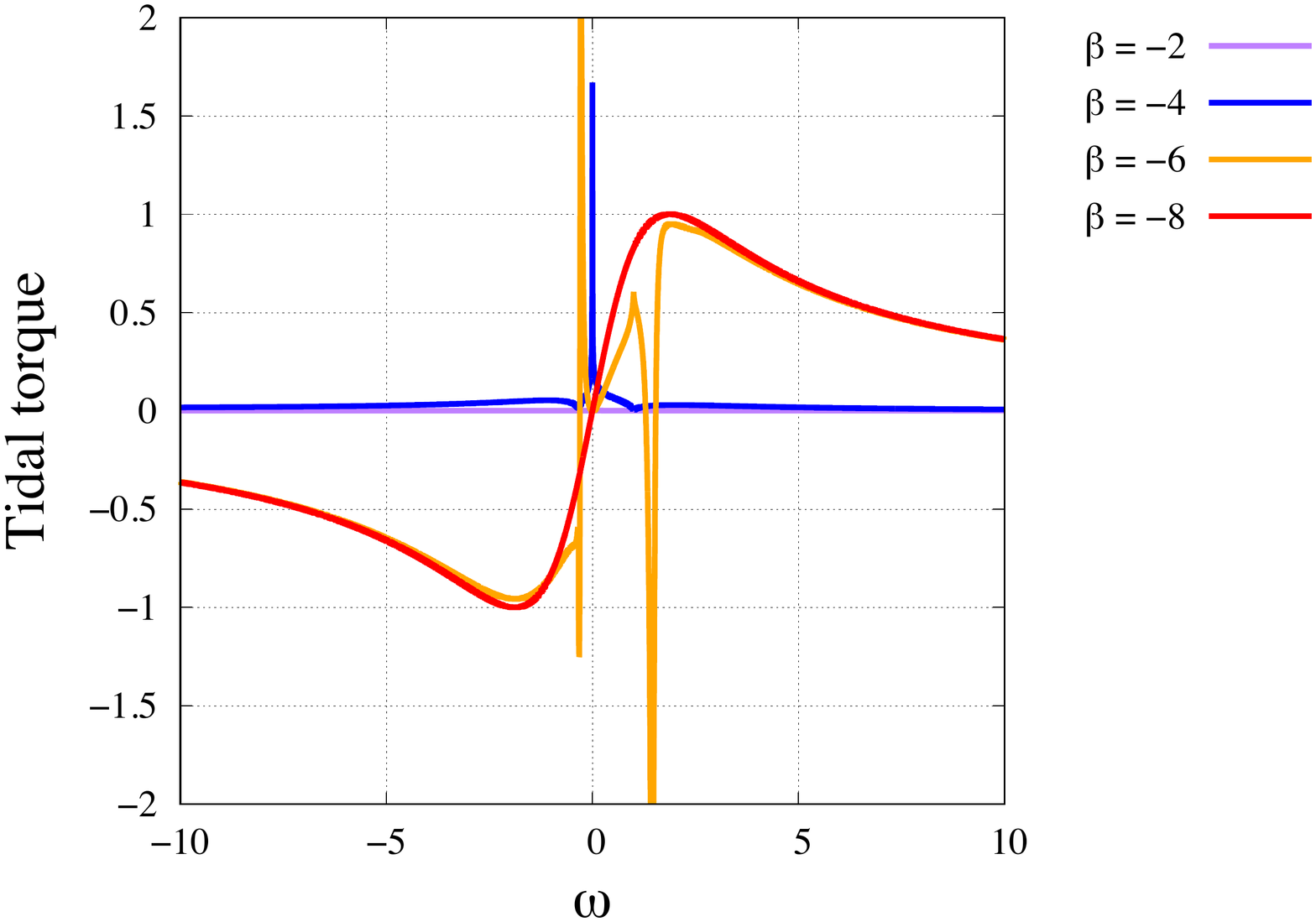} \hspace{0.1cm}  
 \includegraphics[height=0.36\textwidth,trim = 1.5cm 2.5cm 6cm 1.5cm,clip]{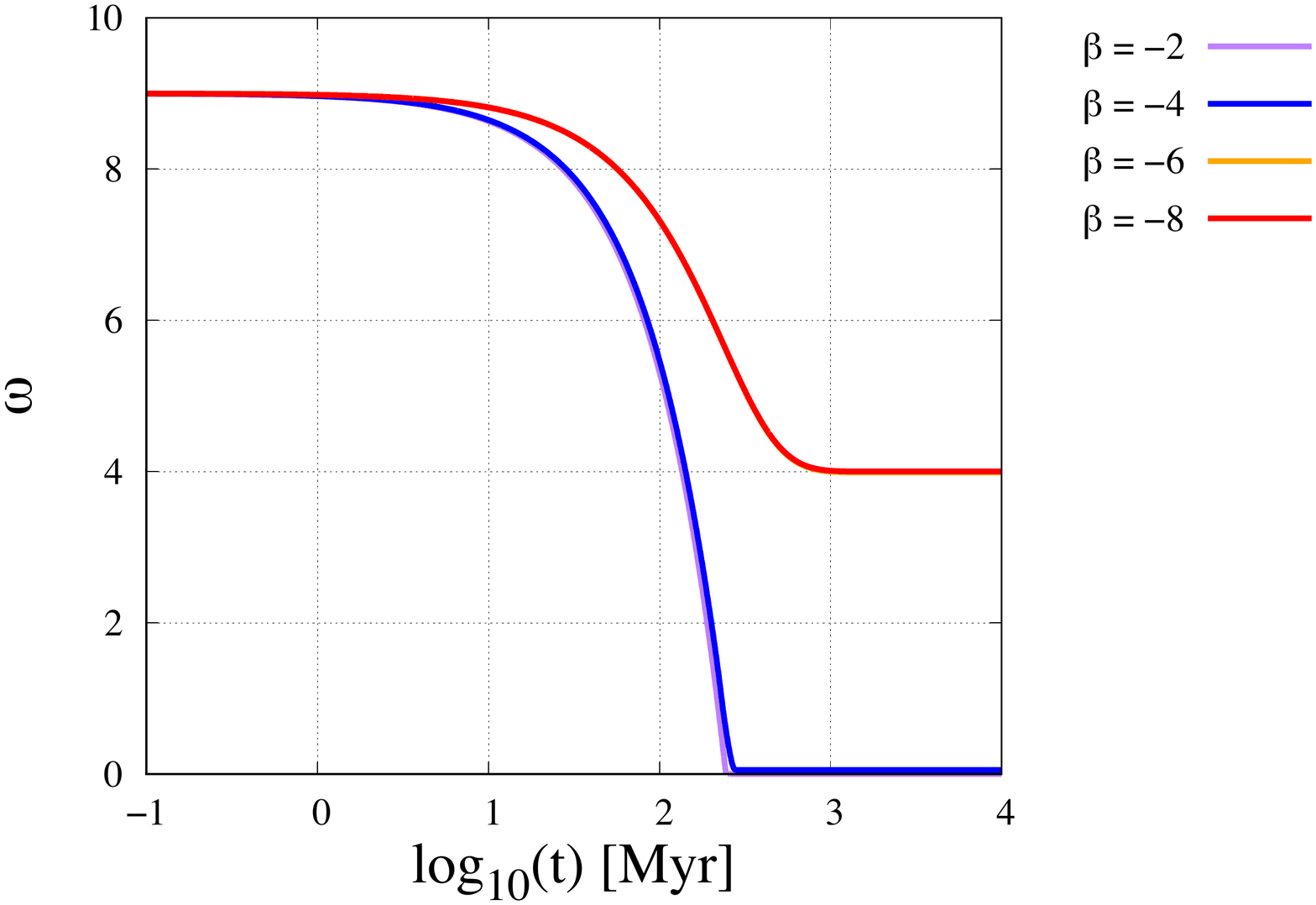}    
 \includegraphics[width=0.12\textwidth,trim = 22cm 3.5cm 0.5cm 1.5cm,clip]{auclair-desrotour_fig3a}    
%% Note the ABSENCE of the extension .pdf  !
  \caption{{\bf Left:} Tidal torque given by the local Cartesian model as a function of the frequency of the perturber $ \omega = \left( \Omega - n_{\rm orb} \right) /n_{\rm norb} $ for various values of the Brunt-V\"ais\"al\"a frequency from weak to strong stable stratification, i.e. \smc{$ \beta = \log \left( N \right) =  \left\{ -8, - 6, -4, -2 \right\} $}. The tidal torque is computed using \smc{Eq.~(\ref{torque_local})}. It is normalized by the maximum value of the torque of the convective case ($ \beta = - 8 $). {\bf Right:} Evolution of the rotation rate of a Venus-like planet for the same Brunt-V\"ais\"al\"a frequencies. The normalized frequency $ \omega = \left( \Omega - n_{\rm orb} \right) / n_{\rm orb} $ is plotted as a function of time (Myr) in logarithmic scale. The value $ \omega = 0 $ corresponds to spin-orbit synchronous rotation. }
  \label{fig:evolution}
\end{figure}

The local model allows us to refine the diagnosis made with the global approach. Considering a planar wave propagating along the longitudinal direction, we obtain for the associated tidal torque the expression

\begin{equation}
\mathcal{T} = 2 \sigma_0  \Im \left\{ \frac{i}{\sigma - i \sigma_0} \left[ 1 - \nu - \frac{\nu}{1 - K \left( \sigma, k_z \right) } \right] \right\},
\label{torque_local}
\end{equation}

\noindent where $ i^2 = -1 $, $ \Im $ \smc{is} the imaginary part of a complex number, $ \nu $ a function of the physical parameters of the system \smc{(forcing frequency, Brunt-V\"ais\"al\"a frequency, spin angular velocity, thermal frequency, colatitude, pressure height scale, horizontal wavenumber)}, and $ K $ a function of the forcing frequency $ \sigma $ and vertical wavenumber $ k_z $ of the associated tidal wave, which characterizes the contribution of internal \smc{gravity} waves to the tidal torque. \smc{On the one hand}, in the low-frequency range ($ \sigma \rightarrow 0 $), $ \nu \rightarrow 0 $ in the case of a neutrally-\smc{stratified} atmosphere, leading to the Maxwell model identified previously. On the other hand, $ \nu \rightarrow 1 $ and $ \left| K \right| \rightarrow + \infty $ in the stably-stratified case, which implies $ \mathcal{T} \approx 0 $. This suggests that the rotation rate of the planet will tend towards a \emph{non-synchronized} state of equilibrium in the neutrally-stratified case and towards the spin-orbit synchronous rotation rate in the stably-stratified one. \smc{This is} illustrated by \smc{Fig.~\ref{fig:evolution}} (right \smc{panel}) where the normalized tidal frequency ($ \omega $) of a Venus-like terrestrial planet is plotted as a function of time for different atmospheric stratifications using \smc{Eq.~(\ref{torque_local})} for the atmosphere and a constant tidal quality factor for the solid part of the planet.The corresponding atmospheric tidal torques are plotted as a function of $ \omega $ (left panel). 

\section{Conclusions}
%%--------------------

Motivated by the understanding of the role played by atmospheric tides in the evolution of the rotation rate of terrestrial planets, we computed the tidal torque generated by the thermally excited semidiurnal tide analytically by using an ab initio linear global model. \smc{We recovered} the behaviour previously obtained with numerical simulations using GCMs, which corresponds to that described by the so-called Maxwell model. \smc{We} showed evidence that the tidal response strongly depends on the structure \smc{and stability of} the atmosphere through its stratification with respect to convection. In a second step, we explored this dependence by using a local Cartesian model, showing that it is determined by the hierarchy of characteristic frequencies of the system. We established the continuous transition between the stably and neutrally stratified cases, which lead to synchronized and non-synchronized rotation states of equilibrium respectively. 

% Optional acknowledgements
% -------------------------
\begin{acknowledgements}
The authors acknowledge funding by the European Research Council through the ERC grants WHIPLASH 679030 and SPIRE 647383, \smc{and the PLATO CNES funding at CEA/IRFU/DAp}.
\end{acknowledgements}

\bibliographystyle{aa}  % A&A bibliography style file (aa.bst)
\bibliography{auclair-desrotour} % your references in file: Yourfile.bib

\end{document}